\documentclass[conference]{IEEEtran}

\ifCLASSINFOpdf
\else
\fi
\newcommand{\ie}{\textit{i.e.,} }
\newcommand{\eg}{\textit{e.g.,} }

\usepackage{fancyvrb}
\usepackage{xcolor}

\definecolor{codered}{HTML}{D9534F}
\definecolor{codegreen}{HTML}{5CB85C}
\definecolor{codeblue}{HTML}{0000FF}
\usepackage{url}
\usepackage{epstopdf}
\usepackage{balance}
\usepackage{color}
\usepackage{listings}
\usepackage{booktabs}
\usepackage{enumitem}
\usepackage{float}
\usepackage{hyperref}
\usepackage[noend]{algpseudocode}
\usepackage{xspace}
\usepackage{listings}
\usepackage{mathtools}
\usepackage{amsmath}
\usepackage{mathrsfs,amsmath} 
\usepackage{soul}
\usepackage[scaled=0.95]{inconsolata}

\usepackage{fancyhdr}

\newcommand{\tool}{\textsc{Elixir}\xspace}

\newcommand{\bugsjar}{Bugs.jar\xspace}
\newcommand{\dfj}{Defects4J\xspace}
\newcommand{\flair}{FLAiR\xspace}
\newcommand{\rexps}{repair-expressions\xspace}
\newcommand{\rexp}{repair-expression\xspace}
\newcommand{\RExps}{Repair-Expressions\xspace}

\newcommand{\baselineOne}{\textsc{Elixir}-Baseline\xspace}
\newcommand{\baselineTwo}{\textsc{Elixir}-NoML\xspace}
\newcommand{\code}[1]{\texttt{#1}}

\IEEEoverridecommandlockouts
\begin{document}
%
\title{\tool: Effective Object-Oriented Program Repair}

\author{\IEEEauthorblockN{Ripon K. Saha,
Yingjun Lyu\IEEEauthorrefmark{1}\thanks{The second author was an intern at Fujitsu Labs of America, CA, USA during the construction of \bugsjar.}, Hiroaki Yoshida,
Mukul R. Prasad}
\IEEEauthorblockA{Fujitsu Laboratories of America, Sunnyvale, CA, USA\\
\IEEEauthorrefmark{1}University of Southern California, Los Angeles, CA, USA\\
rsaha@us.fujitsu.com,
\IEEEauthorrefmark{1}yingjunl@usc.edu,
hyoshida@us.fujitsu.com,
mukul@us.fujitsu.com}}

\maketitle
\thispagestyle{fancy}


\begin{abstract}
This work is motivated by the pervasive use of method invocations in object-oriented (OO) programs, and indeed their prevalence in patches of OO-program bugs. We propose a generate-and-validate repair technique, called \tool\, designed to be able to generate such patches. \tool\ aggressively uses method calls, on par with local variables, fields, or constants, to construct more expressive \rexps, that go into synthesizing patches. The ensuing enlargement of the repair space, on account of the wider use of method calls, is effectively tackled by using a machine-learnt model to rank concrete repairs. The machine-learnt model relies on four features derived from the program context, \ie  the code surrounding the potential repair location, and the bug report. We implement \tool\ and evaluate it on two datasets, the popular \dfj dataset and a new dataset \bugsjar created by us, and against 2 baseline versions of our technique, and 5 other techniques representing the state of the art in program repair. Our evaluation shows that \tool\ is able to increase the number of correctly repaired bugs in \dfj by $85\%$ (from 14 to 26) and by $57\%$ in \bugsjar (from 14 to 22), while also significantly out-performing other state-of-the-art repair techniques including ACS, HD-Repair, NOPOL, PAR, and jGenProg.
\end{abstract}

\section{Introduction}\label{seclbl:intro}

As software applications continue to grow in size and complexity, and fuel the development of new application domains, such as cloud computing, big-data analytics, mobile computing, and software-defined networks, they inevitably produce a corresponding increase in the number of software bugs, and ultimately in the cost of fixing these bugs. For example, a study from the University of Cambridge showed that, as of 2013, the global cost of debugging software  had risen to \$312 billion annually~\cite{bug-cost-study:2013}. Further, the research found that, on average, software developers spend 50\% of their programming time finding and fixing bugs. Automatic software repair techniques have the potential to mitigate some of these costs and thereby increase developer productivity.

Object-oriented (OO) languages dominate the programming landscape today. In fact, 4 of the top 5 languages on the TIOBE Index~\cite{tiobe-index}, namely Java, C++, C\#, and Python, are object-oriented or bear at least some object-oriented features. However, most techniques for automatic program repair~\cite{GenProg:ICSE2012, AE:ASE2013, RSRepair:ICSE2014, SemFix:ICSE2013, DirectFix:ICSE2015, SPR:FSE2015, Prophet:POPL2016, Angelix:ICSE2016, antiPatterns:FSE2016}, with a few notable exceptions~\cite{PAR:ICSE2013, NOPOL:TSE2017, HDRepair:SANER2016, ACS:ICSE2017}, have been developed in the context of C programs, \ie procedural programs. Studies have shown~\cite{Ahsan:ICSEA2009} that bug-patterns can be language specific. Thus, there is a strong need to develop automatic repair techniques targeting object-oriented programs.

One of the core principles of object-oriented language design is the notion of \textit{encapsulation}~\cite{Booch:2004}, whereby the internal data representation of a class (object) and its implementation of operations is hidden from external users of the class. External objects can only access this data and operations through the public methods of the class (object). Thus, the construct of a method invocation (MI) (used inter-changeably with the term \emph{method call} in this paper) on an object, constitutes the basic unit of data access and computation in object-oriented programs. In fact, according to an empirical study we conducted on three popular Java projects (discussed in Section~\ref{sec:empiricalStudy}), as many as
$57\%$ of program statements in each of these applications contain one or more method invocations. Further, according to the same study,  $77\%$  
of {\it one-line} bug-fixes made during the lifetime of each of these projects involved a change to or insertion of a method invocation.  
These data points demonstrate the need to incorporate repair and synthesis of method invocations, in a comprehensive manner, in the repair of bugs in object-oriented programs. For concreteness, the rest of the paper uses Java as a representative of OO-languages. However, the discussion would be equally applicable to other OO-languages, such as C++.

A number of program repair techniques, proposed for Java programs, like PAR~\cite{PAR:ICSE2013}, NOPOL~\cite{NOPOL:TSE2017}, HD-Repair~\cite{HDRepair:SANER2016}, and ACS~\cite{ACS:ICSE2017}, in fact manipulate method invocations in their repairs. However, this is done through very specific schemas and with tight restrictions. For instance, 
NOPOL is the only tool that synthesizes (\ie creates from scratch) method calls, but only on manually specified, side-effect-free, parameter-free method calls and only as guards of inserted if-conditions. 
PAR replaces names or parameters of method calls but only with other names or expressions appearing in other method calls \textit{in the same method}. A plausible reason for such restrictions, documented in a recent work by Martinez and Monperrus~\cite{MartinezM13}, is the combinatorial explosion that would result from expanding the repair space to include a much wider scope of method call modifications and insertions. Figure~\ref{fig:lang-538} shows a simple example of this, using a bug fix from the \emph{Apache Commons Lang} project, that is correctly patched by our proposed tool, \tool. As shown, the patch consists of a single method invocation. However, for constructing such a method invocation, that is correctly typed and in scope, there are more than $800$ concrete candidates! Obviously, a repair approach cannot afford to iterate through each of them.

This work proposes a generate-and-validate repair technique, called \tool, developed for the repair of object-oriented programs. A key innovation in \tool\ is the aggressive use of method calls, on par with local variables, fields, and constants, to construct more expressive \textit{repair expressions}, that go into synthesizing patches. The ensuing enlargement of the repair space is effectively tackled by using a machine-learned model to rank concrete repairs. The machine-learned model relies on four features derived from the \textit{repair context}, \ie  the code surrounding the potential repair location, and from the bug report. The features describe a given identifier, which could represent a local variable or object, a constant, or a method call. In particular the features quantify (1) how frequently the identifier has been used in the current context, (2) the distance of the place of last use from the repair location, (3) whether ``similarly named" tokens have been used in the repair context, and (4) if the identifier or sub-tokens thereof have been referenced in the bug report (if one is present). Variants of some of these features have been used in heuristics employed in code recommendation~\cite{Asaduzzaman:2014, Nguyen:2016}, bug localization~\cite{zhou2012should, ye2014learning}, and program repair~\cite{liu2013r2fix}. However, a key contribution of our work is the choice and specific incarnation of those features in the present context, and their use in a machine-learned model used to guide a program repair technique. Also novel is the insight that such a technique can effectively navigate a huge repair space to fix bugs involving method invocations in OO-programs, specifically Java.

We implement and evaluate \tool\ on two datasets, the popular Defects4J dataset and a new dataset Bugs.jar created by us, and against two baseline versions of our technique, as well as five other tools/techniques representing the state of the art in program repair. Our evaluation shows that \tool\ is able to increase the number of correctly repaired bugs in \dfj by $85\%$ (from 14 to 26) and by $57\%$ for \bugsjar (from 14 to 22), while also significantly out-performing other state-of-the-art repair techniques including ACS~\cite{ACS:ICSE2017}, HD-Repair~\cite{HDRepair:SANER2016}, NOPOL~\cite{NOPOL:TSE2017}, PAR~\cite{PAR:ICSE2013}, and jGenProg~\cite{martinez2016automatic}.
This paper makes the following key contributions:
\begin{itemize}
	\item \textbf{Empirical study:} An empirical study highlighting the prevalence of method invocations in patches of Java programs, as a motivation for our proposed technique.
	\item \textbf{Technique:} A novel technique, \tool\, that aggressively employs method invocations in constructing repairs for Java programs, and object-oriented programs in general.
	\item \textbf{Implementation:} An implementation of \tool\ in our in-house Java repair framework, along with two baseline versions of \tool.
	\item \textbf{Dataset:} A new, large dataset of 1,158 bugs and patches,  \bugsjar, made available to the research community at \cite{bugsdotjar}, to complement existing datasets like \dfj .
	\item \textbf{Evaluation:} A comprehensive evaluation of \tool\ on two datasets, Defects4J and \bugsjar, and against seven competing techniques, including two baseline versions of \tool\ and five external tools/techniques, ACS, HD-Repair, NOPOL, PAR, and jGenProg.
\end{itemize}

\section{Motivation}\label{seclab:motivation}

This section demonstrates the prevalence of method invocations (MI) and  MI-related bugs, in Java programs, through an empirical study and two real-world motivating examples.

\subsection{An Empirical Study on the Method Invocation Construct and Its Relevance to Bugs in Java Programs}
\label{sec:empiricalStudy}

The construct of method invocation (MI) is fundamental to orchestrating data access and computation in OO-programs, and indeed for enforcing key OO features, such as encapsulation. However, it is natural to ask if real-world OO-programs (\eg Java applications) demonstrate a \emph{quantitatively greater use} of MIs than procedure-oriented programs (\eg C). 
To investigate this empirically, we selected three Java projects: Eclipse JDT,  Platform, and BIRT, which are popular Java projects used in bug localization research~\cite{ye2014learning} and the ManyBugs~\cite{le2015manybugs} benchmarks, widely used in C program repair research, representing C applications. The Java projects were intentionally chosen to be distinct from our experimental datasets (\dfj and \bugsjar), to guard against learning bias.
We parsed all the source code files of both Java and C applications, excluding test cases, and counted the fraction of executable statements having an MI or a function call. Our results show that, on average, 57\% of statements in each of the Java applications have an MI, compared to only 33\% for the C programs. Thus, this study, while decidedly limited in scope, supports the hypothesis that Java programs use MIs substantially more than C programs.

\begin{figure}[t]
\begin{center}
\includegraphics[width=0.9\linewidth]{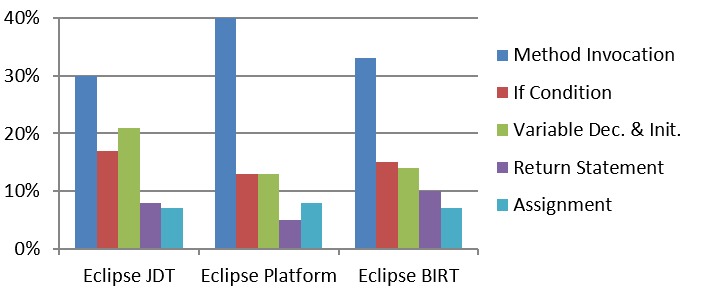}
\end{center}
\vspace{-15pt}
\caption{Distribution of Bug-Fixing Changes. Based on 1186, 1031, and 985 one-line bug fixes in Eclipse JDT, Platform, and BIRT projects respectively.}
\label{fig:causes}
\vspace{-15pt}
\end{figure}

We further analyzed all the one-line bug-fixes in each of the three Java projects to investigate how often MIs appear in those patches. We focus on one-line bug-fixes since current automatic program repair tools mainly target such fixes. We used ChangeDistiller~\cite{Fluri:2007} to extract the one-line bug-fixes throughout each project's history, and automatically classified each patch into one of a few \emph{mutually exclusive} categories based on the type of that statement, such as method invocation, if condition, variable declarations and initializations, return statements, and assignments \textit{etc}. Figure~\ref{fig:causes} plots this classification, per project, for the top 5 categories. The results show that 30\%-40\% of one-line bug-fixes, the most dominant class, are stand-alone MI statements. And this does not include indirect MI changes, for example, changing an MI in the guard of an {if condition}, currently classified as an {\it if condition} change in Figure~\ref{fig:causes}. Manually investigating each of the non-MI labeled bug-fixes revealed that almost 60\% of {\it if condition changes}, and at least 80\% of other changes (variable initializations, assignments, and return expressions) involved MIs. In aggregate, 77\% of the studied one-line bug-fixes involved MI changes, either stand-alone or part of another construct.

These results demonstrate the need to incorporate MI-related modifications, in a comprehensive manner, in the search space examined by a repair tool. The following examples illustrate the shortcomings of current tools in this respect.

\subsection{Motivating Examples}
Now we discuss two real-world examples that are beyond the scope of current repair tools, since they entail synthesis of substantially new MIs, typically rendered infeasible by a combinatorial explosion in the number of candidate patches. 

\begin{figure}[t]
\fbox{
\begin{minipage}{0.945\linewidth}\scriptsize\it
\setlist[description]{labelindent=0mm,style=multiline,leftmargin=26mm,labelsep=0mm,itemindent=0mm}
\begin{description}
\item[Bug Report Summary:] DateFormatUtils.format does not correctly change
                           Calendar TimeZone in certain situations
\end{description}
\end{minipage}
}
\begin{Verbatim}[frame=single,fontsize=\scriptsize,commandchars=&\[\]]
     public StringBuffer format(Calendar calendar, StringBuffer buf) {
         if (mTimeZoneForced) {
&fvtextcolor[codeblue][+            calendar.getTime();] /// LANG-538
             calendar = (Calendar) calendar.clone();
             calendar.setTimeZone(mTimeZone);
         }
         return applyRules(calendar, buf);
     }
\end{Verbatim}
\vspace{-10pt}
\caption{The bug report summary (top) and fix (bottom) for \code{LANG-538}}
\label{fig:lang-538}
\vspace{-3mm}
\end{figure}

\begin{figure}[t]
\fbox{
\begin{minipage}{0.945\linewidth}\scriptsize\it
\setlist[description]{labelindent=0mm,style=multiline,leftmargin=26mm,labelsep=0mm,itemindent=0mm}
\begin{description}
\item[Bug Report Summary:] Field not initialized in constructor:\\
org.apache.commons.lang.LocaleUtils.cAvailableLocaleSet
\end{description}
\end{minipage}
}
\begin{Verbatim}[frame=single,fontsize=\scriptsize,commandchars=&\[\]]
     public static boolean isAvailableLocale(Locale locale) {
&fvtextcolor[codered][-       return cAvailableLocaleSet.contains(locale);]
&fvtextcolor[codeblue][+       return availableLocaleList().contains(locale);]
     }
\end{Verbatim}
\vspace{-3mm}
\caption{The bug report summary (top) and fix (bottom) for \code{LANG-304}}
\label{fig:lang-304}
\vspace{-15pt}
\end{figure}

Figure~\ref{fig:lang-538} presents a bug-fix (Bug ID: \code{LANG-538}) in the Apache Commons Lang project, taken from the popular \dfj dataset.
This patch requires the insertion of a new method invocation statement and is outside the repair space of current repair tools, since including such MIs would increase the repair search space significantly. For example, to fix this bug, \tool synthesizes 836 valid MIs for that location (Table~\ref{tbl:candidate:patches}). Certainly, validating such a large number of candidates, for a given repair location and a transformation schema, in a brute-force fashion is not practical. The only way the existing tools (such as SPR~\cite{SPR:FSE2015}, PAR~\cite{PAR:ICSE2013}, or GenProg~\cite{GenProg:ICSE2012}) could attempt this bug fix would be by copying and pasting the same exact statement \code{calendar.getTime()} from elsewhere in the code. However, this statement is not present elsewhere.

Figure~\ref{fig:lang-304} shows another bug (\code{LANG-304}) also from Commons Lang in \dfj.
From an automatic repair point of view, this is also a non-trivial fix since the object of an MI is replaced by another MI that returns a compatible (type \code{List}) but not exactly the same type (type \code{Set}) of object. Current repair tools do not include such complex MI transformations in their repair space, to keep the search manageable. Also, the patch cannot be copied verbatim from elsewhere in the program either.

\textbf{Proposed approach:} Our proposed technique, \tool can synthesize the correct patches for both the above bugs by, (1) first synthesizing a population of candidate patches and, (2) then ranking this candidate population and validating only the top several candidates.
To synthesize the candidates \tool first extracts atomic elements such as objects, variables, and literals in scope, and then synthesizes valid MI (\eg \code{obj.foo(a,b)}) and field access (\eg \code{obj.a}) expressions. These expressions constitute the building blocks (termed \rexp) for synthesizing patches, and are then plugged into various program transformation schemas to generate candidate patches.

To effectively deal with the large set of candidate patches resulting from a rich set of \rexps, \tool ranks the candidates using a machine-learned model and only selects the top few for validation against the test-suite. The machine-learned model is built on a set of four simple, but potent features, described in Section~\ref{subsec:ranking}. 
For our motivating example, bug \code{LANG-538}, when \tool instantiates the MI-insertion schema (described in Section~\ref{lb:schemas}), the correct patch is ranked at $7^{th}$ out of $836$ candidate patches (Table~\ref{tbl:candidate:patches}) and hence can be quickly validated through the test-suite.

\begin{table}[t]
\vspace{-1pt}
\setlength{\tabcolsep}{4pt}
\begin{center}
\caption{Candidate Patches for LANG-538}
\vspace{-5pt}
\label{tbl:candidate:patches} 
\begin{tabular}{cl}
\toprule
{\bf Rank}&{\bf Synthesized MIs}\\\midrule
1& format(calendar)\\
2& calendar.clear()\\
3& format(calendar, buf)\\
:&\\
6& calendar.setLenient(mTimeZoneForced)\\
7& {\bf calendar.getTime()}\\
:&Other 800+ Candidates\\
\bottomrule
\end{tabular}
\end{center} 
\vspace{-15pt}
\end{table}



\section{\tool}\label{seclab:techniques}

\begin{figure*}
\begin{center}
\includegraphics[width=0.9\textwidth]{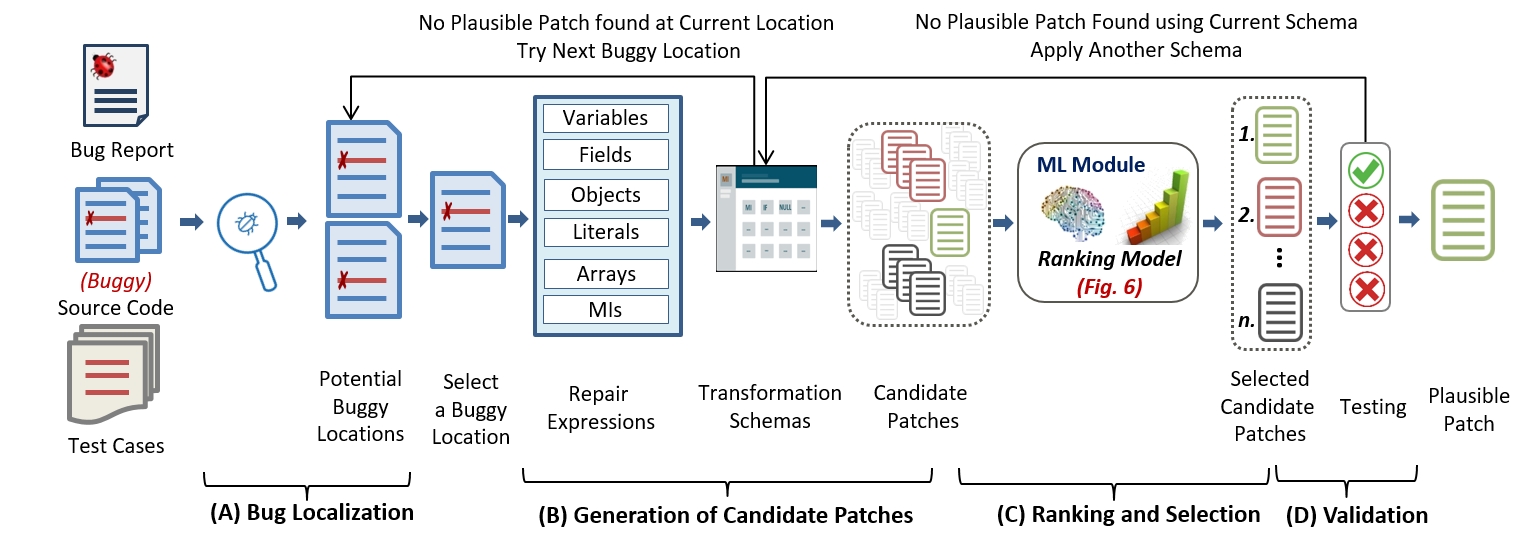}
\end{center}
\vspace{-20pt}
\caption{Overview of \tool.}
\vspace{-15pt}
\label{fig:elixir}
\end{figure*}

The overall structure of \tool is presented in Figure~\ref{fig:elixir}. For a given bug, \tool takes as input a buggy program, a test suite (having at least one bug reproducing test case), and optionally a bug report, and produces a patch  that passes all the test cases, \ie fixes the bug. \tool works in four steps. 

\begin{description}
\item [(A)] {\bf Bug Localization.} This step identifies a list of suspicious statements in the buggy program. Then for each potential buggy statement (repair location), \tool performs the following steps until a {\it plausible patch}\footnote{A plausible patch is one that simply passes all test-cases (including failing tests) in the test suite, but may still be incorrect because the test-suite may provide an incomplete specification.} is found.
\item [(B)]{\bf Generating Candidate Patches.} \tool includes a set of program transformation schemas described in Section~\ref{lb:schemas}. For each schema, \tool generates a list of candidate patches by plugging in various \rexps into the schema, and performs the following steps until a plausible patch is found. 
\item [(C)]{\bf Ranking and Selection of Candidate Patches.} \tool uses a machine-learned model to rank the candidate patches and selects the top $N$ patches for validation.
\item [(D)]{\bf Validating Selected Candidate Patches.} \tool applies the selected patches one at a time, beginning from the top of the ranked-list, on the buggy program, and runs the test cases. If all the test cases pass, \tool terminates and returns that patch as a plausible patch.
\end{description}


\vspace{-5pt}
\subsection{Bug Localization} \label{lbl:bugloc}
\tool uses the Ochiai technique~\cite{Abreu:2006}, a popular existing spectrum-based bug localization approach to identify potential buggy statements. According to the Ochiai technique, \tool instruments the program at statement level, and collects test spectrum\textendash \ie the statements executed by each test case, and computes a suspiciousness score for each statement. Finally, all statements are ranked in a descending order of supiciousness score, \ie the top statement is the most suspicious.

\vspace{-5pt}
\subsection{Generation of Candidate Patches}

Fixing a bug involves applying appropriate changes at the buggy location. Allowing arbitrarily complex transformations can results in an infinite number of candidate patches. Therefore, program repair tools typically define their repair space through a fixed set of parameterized program transformation schemas, paired with a restricted set of expressions to instantiate those schemas. We term these expressions as \emph{\rexps}. For example, the schema \emph{Insertion of a Method Invocation} instantiated with the MI \rexp \code{calendar.getTime()} yields the patch in Figure~\ref{fig:lang-538}. 

\subsubsection{Program Transformation Schemas} \label{lb:schemas}
\tool applies the following program transformation schemas in the presented order to produce candidate patches for a given statement. 

\begin{description}
\item [(T1)] \textbf{Widening Type: } For a variable declaration statement, this schema replaces the type of the variable with a widened type -- \eg \code{float} to \code{double}.
\item [(T2)] \textbf{Changing Expression in Return Statement:} This schema replaces a returned expression by another expression having compatible types.
\item [(T3)]\textbf{Checking Null Pointer:} If a statement has an object reference, this schema adds an {\it if guard} that ensures no null object is accessed.
\item [(T4)]\textbf{Checking Array Range and Collection Size:} If a statement has array references or collection objects, this schema adds an {\it if guard} to ensure that all array or collection accesses are within range, to prevent exceptions.
\item [(T5)]\textbf{Changing Infix Boolean Operator:} This schema includes common mutation operators from mutation testing research. For example, an infix expression like $a>b$ can be changed to $a \ge b$, $a<b$, and so on.
\item [(T6)]\textbf{Loosening and Tightening Boolean Expression :} If a \code{boolean} expression is an if condition or in a return statement, this schema may remove or add predicates.

\item [(T7)]\textbf{Changing Method Invocation (MI):} This is a complex schema comprised of the following schemas:
\begin{itemize}
\item{\bf Replacing Object Expression:} Replaces the object reference by another compatible-typed expression.
\item{\bf Replacing Method Name:} Replaces the method name with another method name having the same signature.
\item{\bf Replacing Argument: }Replaces an argument expression by another expression having compatible types.
\item{\bf Replacing a full MI by a synthesized MI:} Replaces the complete MI by a synthesized MI (can also be an overloaded MI) that returns a compatible type.
\end{itemize}
\item [(T8)]{\bf Insertion of a Method Invocation:} This is a new schema in \tool. It synthesizes MIs, and inserts them as a part of an expression or as a complete statement.
\end{description} 

\subsubsection{Synthesis of \RExps}
One of the key features of \tool is that it uses a rich set of \rexps that are effective in fixing real bugs in OO-programs. In particular, this entails a generous use of MIs and object accesses. Figure~\ref{fig:ingredient} shows the grammar of \rexps used in \tool. 

\begin{figure}
\footnotesize
\begin{center}
\begin{minipage}{\linewidth}\scriptsize
\begin{align*}
    literal &\rightarrow {\bf boolean}\ |\ {\bf number}\ |\ {\bf null}\\
    varable &\rightarrow {\bf id}\\
    field &\rightarrow {\bf id}.{\bf id}\\
    array &\rightarrow {\bf id}[expression]\\
    expression &\rightarrow literal\ |\ variable\ |\ field\ |\ array\\
    argumentList &\rightarrow argumentList, expression\ |\ expression\\
    methodInvocation &\rightarrow {\bf id}(argumentList)\ |\ {\bf id}.{\bf id}(argumentList)
\end{align*}
\end{minipage}
\end{center}
\vspace{-2mm}
\caption{Grammar to Describe the Specific \RExps in \tool. It should be that this grammar simply presents the structures of expressions. Please refer to the relevant documentation~\cite{Java:specification} for the accurate grammar.}
\vspace{-10pt}
\label{fig:ingredient}
\end{figure}

The synthesis of \rexps involves i) extracting relevant program elements in scope, and ii) creating \rexps using them. Specifically, given a repair location, \tool extracts all the local variables and literals in scope, fields in the same class, and all the public fields in other classes that are relevant to the buggy class. Here, relevant classes are ones whose fields are accessed or methods are invoked within the buggy method. For example, for the bug in Figure~\ref{fig:lang-538}, \code{Calendar} and \code{StringBuffer} constitute relevant classes. Further, \tool extracts the signatures of all the methods that can be invoked from the repair location. Next, \tool creates field expressions (\eg \code{a.b}), and concrete MIs using the extracted  method signatures, variables, literals, and fields, as per Figure~\ref{fig:ingredient}. All these variables, literals, fields, and concrete method invocations constitute the pool of \rexps used in the next step.

\subsubsection{Synthesis of Candidate Patches}

The synthesized \rexps are used to instantiate the program transformation schemas to produce a set of concrete candidate patches. 
It should be noted that the first five schemas of \tool (\textbf{T1} - \textbf{T5}) require only a small set of \rexps. For example, we have to only add null checkers for only those objects that are accessed in the repair location. Using widening type schema we can change {\tt int a;} to {\tt \{long | float | double\} a;}. Therefore, these schemas generate a small set of concrete patches. However, the rest of the schemas involve any \rexps defined in Figure~\ref{fig:ingredient}. Therefore, they may produce a significantly large number of candidate patches. Although generation of patches is fast, compiling the program after applying a patch, and testing it is expensive. To cope with this search space explosion, \tool ranks all the candidate patches and selects the top $N$ patches for validation.

\subsection{Ranking and Selection of Candidate Patches}
\label{subsec:ranking}

Ranking candidate patches is a challenging problem since any valid (\ie compilable) patch can be the correct patch. Our key insight is that the program context and the bug report may provide valuable clues to identify the truly \emph{relevant} patches. Therefore, we propose a machine learning technique to rank and select candidate patches. From a machine learning standpoint, the selection of repair candidates can be viewed as a binary classification problem, where our objective is to determine whether a particular candidate patch is relevant to a particular program context. Furthermore, we can compute a relevance score of each candidate patch, and use that to rank all the candidate patches. To this end, we use logistic regression, which is a widely used machine learning technique in practice.

\begin{figure}
\begin{center}
\includegraphics[width=\linewidth]{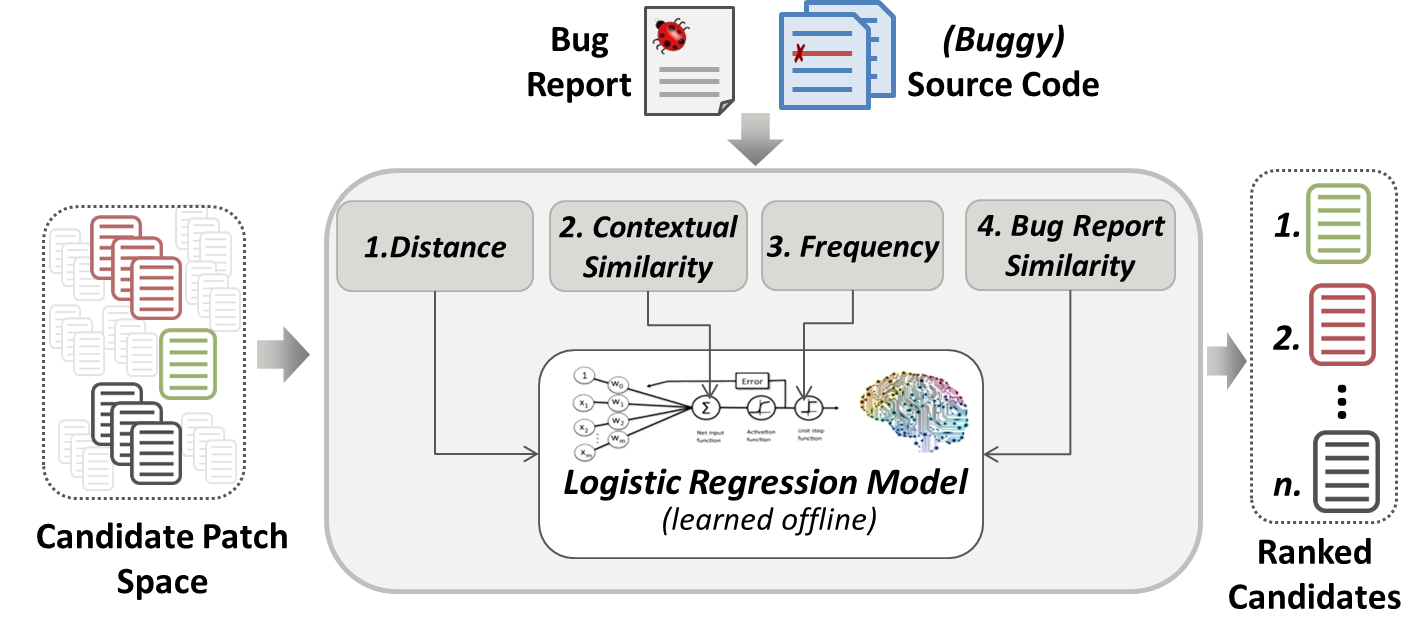}
\end{center}
\vspace{-20pt}
\caption{Ranking and Selection of \RExps.}
\vspace{-15pt}
\label{fig:patch-sel}
\end{figure}

Figure~\ref{fig:patch-sel} presents the overall approach for ranking candidate patches. Given a set of candidate patches, \tool first extracts four feature scores for each candidate patch from the used \rexps in it. These feature scores are calculated based on the repair context and the bug report. Then these feature scores are passed to an already trained logistic regression model that computes a probability score for each candidate patch representing its relevance. The learned logistic regression model is trained offline in advance using the same features from a set of previous bug fixes (training dataset). The subsequent sections describe the approach in more detail.

\subsubsection{Selection of Feature Set and Calculation of Feature Scores}

Based on an extensive study on prior literature on code completion, bug localization, and program repair techniques, we selected four features  for our task: i) distance, ii) contextual similarity, iii) frequency in the context, and iv) bug report similarity. Note that \rexps used in a candidate patch is composed of one or more elements (e.g., variables/fields/literals). A variable itself is a single-element \rexp, whereas an MI is multi-element \rexp since several variables and objects may involve there. Therefore, we compute the feature scores of a patch in terms of the feature scores of new \rexps in the patch. For example, when we change a method parameter, the feature scores of the new parameter represent the patch feature scores. Therefore, \tool first computes the feature score at element-level and then at patch-level.

\textbf{Distance.}  Our first insight is that the more a \rexp is composed of closer elements to the repair location, the more it is relevant. PAR~\cite{PAR:ICSE2013} uses a similar concept in terms of AST nodes to sort the candidate variables. To compute the distance score of a \rexp, \tool first computes the distance score of each element ($\xi$) in the \rexp from the repair location ($\Re$) using Equation~\ref{dist}.

\begin{equation}
\label{dist}
 \mathscr{S}_d = 
  \begin{cases} 
   1 - \frac{ld(\xi,\Re)}{len(m)} & \text{if\space} len(m) \ge ld(\xi,\Re) \\
   0       & \text{otherwise}
  \end{cases}
\end{equation}

where $ld(\xi,\Re)$ is the number of comment-free lines  between $\Re$ and the closest occurrence of $\xi$, and $len(m)$ is the number of comment-free lines in the method. For multi-element \rexp, we average the distance scores of elements.

\textbf{Contextual Similarity.} Repair context, i.e., the surrounding code of repair location often provide useful hint to determine which \rexps are more consistent than others. Program context has been effectively used in auto code completion~\cite{Hindle:2012}, API~\cite{Nguyen:2016} and parameter recommendation~\cite{Asaduzzaman:2014}. Our insight is that the more a \rexp is textually similar to the context, the more relevant it is for the repair location. We set the size of program context to six lines, i.e., three lines before and after the repair location. This size has been found to be effective in a recent work on API recommendation~\cite{Nguyen:2016}. We compute the contextual similarity between a \rexp and its program context as follows: 

\begin{enumerate}
\item We extract the identifier names of the given \rexp, and split CamelCase identifiers ({\tt cAvailableLocaleSet}) into a set of tokens ({\tt c}, {\tt available}, {\tt locale}, and {\tt set}), which we call ($S1$).

\item We extract all the identifier names from the context. If the CamelCase \rexp identifiers are exactly present in the context, we remove it. Otherwise, it would get similarity with itself.
We split the resulting context identifiers, which becomes the context token set $S2$.

\item We compute the token similarity using Jaccard Similarity Coefficient (Equation~\ref{tokensim})
\vspace{-5pt}
\begin{equation}
\label{tokensim}
\mathscr{S}_{context} = \frac{|S1 \cap S2|}{S1 \cup S2}
\end{equation}
\end{enumerate}

\textbf{Frequency in the context.} Our third insight is that the more a \rexp is composed of frequently used elements, the more it is relevant in that context. However, it depends on the type of the element. For example, an object or a variable may be used repeatedly in a program context to perform some operations on it. However, an MI may not be used repeatedly in the same context. We performed a ``Correlation-based Feature Subset Selection" technique from Weka toolkit~\cite{Hall1998} on our training dataset and found that frequency is not correlated with choosing method names but correlated with variables and objects. Therefore, for multi-element \rexp we average the frequency of only objects and variables.

\textbf{Bug Report Similarity.} Bug reports have been widely used in information retrieval based bug localization~\cite{zhou2012should, ye2014learning, Ye:2016}. The idea is based on the fact that bug reporters use similar words to describe a bug that are in the buggy source code. Sometimes, reporters also  write about possible fixes. Liu et al.~\cite{liu2013r2fix} used bug reports for fixing buffer overflows, null pointer bugs,
and  memory  leaks in C programs. Our insight is that the information from the bug report can be used in prioritizing \rexps. To this end, we calculate the similarity score of a \rexp with bug report in terms of token similarity using Equation~\ref{tokensim}, where $S1$ and $S2$ are the set of \rexp  and the bug report tokens respectively.

\subsubsection{Logistic Regression Model for Ranking}

The objective of our approach is to learn a model that aggregates the feature scores of a candidate patch ($\mathcal{P}$) in such a way that, for a given repair location, the candidate patch that are highly relevant would get a higher score than the irrelevant patches.
Let us assume that for a given candidate patch, $\mathcal{P}$, the four feature scores: distance, bug report similarity, context similarity, and frequency are $\mathscr{S}_{dist}$, $\mathscr{S}_{con}$, $\mathscr{S}_{br}$, and $\mathscr{S}_{frq}$.
\tool aggregates the feature scores by a weighted sum:
\vspace{-2pt}
\begin{equation}
\label{lbl:aggregator}
f(\mathcal{P},\theta) = \alpha \times \mathscr{S}_{dist} + \beta \times \mathscr{S}_{con} + \gamma \times \mathscr{S}_{br} + \zeta \times \mathscr{S}_{freq}
\end{equation}

Here $\theta$ is the weight vector $[\alpha, \beta, \gamma, \zeta]$. These weights ($\theta$) are learned from a training dataset.

For a given patch $\mathscr{P}$ and its feature vector $[\mathscr{S}_{dist}, \mathscr{S}_{con}, \mathscr{S}_{br}, \mathscr{S}_{freq}]$, our objective is to compute the probability of $\mathcal{P}$ being relevant to the program context. Logistic regression machine learning technique is a powerful statistical way of modeling a binomial outcome with a probability score. Logistic regression function is defined as:

\vspace{-10pt}
\begin{equation}
\label{logistic}
\sigma(t) = \frac{1}{1 + e^{-t}}
\end{equation}
\vspace{-10pt}

Substituting $t$ in Equation~\ref{logistic} by $f(\mathcal{P},\theta)$, we get:

\vspace{-10pt}
\begin{equation}
\label{substitued}
\sigma(f(\mathcal{P},\theta)) = \frac{1}{1 + e^{-f(\mathcal{P},\theta)}}
\end{equation}

For binomial classification, we can assume that a data instance follows Bernoulli distribution~\cite{AML}, which is:

\vspace{-10pt}
\begin{equation}
\label{pdf}
p(f(\mathcal{P},\theta),y|\theta) = \sigma(f(\mathcal{P},\theta))^y(1-\sigma(f(\mathcal{P},\theta)))^{(1-y)}
\end{equation}
\vspace{-10pt}

where y is 1 when the candidate patch is relevant and 0 when it is irrelevant. Therefore, to learn the weights of the model, $\theta$, we compute $p(\theta|\xi,y)$, which is the posterior probability, from a training dataset. We use Weka's implementation of logistic regression~\cite{leCessie1992} to learn $\theta$.




\section{Experimental Setup}\label{setup}

\subsection{Implementation} \label{lbl:implementation}
{\bf \tool} is implemented on the top of an automatic program repair framework, called \flair that we developed. \flair has its own bug localization system, various program transformation schemas, an in-memory compilation system, JUnit test case execution system, and a run-time data monitoring system. All the tools in \flair are written in Java, leveraging existing libraries where possible. More specifically, FLAIR bug localization system uses the ASM byte code library~\cite{asm} to instrument programs' source code. \flair  uses Spoon~\cite{pawlak:hal-01169705} library to modify a program at the abstract syntax tree (AST) level. After applying a repair schema, FLAIR leverages javax.tools~\cite{javax} for in-memory compilation. Then FLAIR uses JUnit APIs to run the test cases programmatically. We also implement two other variants of \tool on \flair framework.

{\bf \baselineOne} uses exactly the same program transformation schema as in \tool. However, it uses the \rexps following the existing tools such as ACS, PAR, and HD-Repair. This baseline helps us to demonstrate the contribution of our rich set of \rexps.

{\bf \baselineTwo} uses both \tool's schemas and \rexps. However, it selects N patches randomly instead of any machine learning technique. This baseline helps us to demonstrate the contribution of our proposed ranking model.
\vspace{-5pt}
\subsection{Dataset} 
In order to rigorously evaluate \tool, we used two dataset.

\subsubsection{\dfj} Our first dataset is the popular \dfj dataset~\cite{just2014defects4j}. We used the same four subjects from \dfj that the existing repair tools are evaluated on. Table~\ref{tbl:dfj} (taken from~\cite{just2014defects4j}) presents the details of the dataset.

\begin{table}[t]
\vspace{-1pt}
\setlength{\tabcolsep}{4pt}
\begin{center}
\caption{Details of Subjects in \dfj~\cite{just2014defects4j}}
\vspace{-5pt}
\label{tbl:dfj} 
\begin{tabular}{lccc}
\toprule
{\bf Subject}&{\bf \#Bugs}&{\bf KLOC}&{\bf \#Tests}\\\midrule
Commons Math&106&85&3,602\\
Commons Lang&65&22&2,245\\
Joda-Time&27&28&4,130\\
JFreeChart&26&96&2,205\\
\bottomrule
\end{tabular}
\end{center} 
\vspace{-20pt}
\end{table}

\begin{table}
\setlength{\tabcolsep}{2pt}
\begin{center}
\caption{Details of \bugsjar}
\vspace{-10pt}
\label{tab:bugsjar}
\begin{tabular}[h]{@{}lcrrrr@{}}
\toprule
{\bf Project} &{\bf Tags By} & {\bf Commits} & {\bf Bugs} &  {\bf Size} & \textbf{Bugs} \\
 & {\bf Apache}& &{\bf Reports} & [KLoC] & \textbf{Selected} \\
\midrule
Accumulo & database& 8,714 & 2,041 & 458 & \textbf{98} \\
Camel & network-client/server& 24,096 & 1,081 & 257 & \textbf{147} \\
Commons Math & library & 5,994 & 635 & 187 & \textbf{147} \\
Flink  & big data& 8,906 & 2,070 & 345 & \textbf{70} \\
Jackrabbit Oak & XML &10,810 & 1,686 & 228 & \textbf{278} \\
Log4J2  &library & 6,971& 784 & 104 & \textbf{81} \\
Maven & build management & 10,264 & 2,863 & 100 & \textbf{48} \\
Wicket & web framework& 19,386&3,770 & 177 & \textbf{289} \\
\midrule
\textbf{Total} && 95,141 & 14,930 & 1,856 & \textbf{1,158} \\
\bottomrule
\end{tabular}
\end{center}
\vspace{-25pt}
\end{table}

\subsubsection{\bugsjar} Our second dataset is \bugsjar, created by us. The reasons for creating \bugsjar{} are twofold: i) since \tool is an ML-based approach, we need a training dataset, ii) evaluating \tool on a different dataset than \dfj.

\bugsjar is a large-scale full-fledged real-world bug dataset. Since this is a new dataset, we briefly discuss our methodology to create \bugsjar. Each bug in \bugsjar contains (1) the buggy version of the source code, (2) a test-suite, serving as a correctness specification, comprising at least one failing (bug reproducing) test case and one passing test case (to guard against regression), and (3) the developer's patch to fix the bug, which passes all the test cases. Furthermore, each bug in \bugsjar has the original bug report associated with it. 

{\bf Design Criteria.} The design of our dataset was driven by four broad criteria: i) real-world relevance: having large, active projects, with a rich development history, ii) diversity: having projects covering the spectrum of applications, iii) reproducibility: having consistently reproducible bugs, and iv) automatability: building and testing the projects automatically.

{\bf Methodology.} After conducting a rigorous search on GitHub and Google Code, we found that projects developed by the Apache Foundation fulfill our real-world relevance and automatability criteria. Further, there are several hundred projects in this ecosystem and projects are tagged with one or more of 28 different keywords, such as library, big data, etc., representing the application domain of the project. Thus we chose the Apache ecosystem on GitHub, which has 260 such tagged projects, for constructing our dataset. Then we selected the top 8 groups, each of which has at least 20 projects. This strategy respects our diversity criterion.
Then for each group, we selected a representative subject that has at least 50KLoc and 5000 commits. Then for each project, we identified the bug fixing commits following Apache developers' convention that a Bug ID is present in the commit message. Then we consistently reproduce bugs by running test cases at least 10 times. Finally, we manually verified each reproducible bug.  Table~\ref{tab:bugsjar} provides the details of \bugsjar.

\subsection{Research Questions}
We evaluate \tool with respect to four research questions:
\begin{description}

\item [RQ1:]How effective \tool is compared to state-of-the-art on \dfj?

\item [RQ2:]What is the contribution of \rexps, and ranking and selection model of \tool?

\item [RQ3:]Are all features used in \tool contributed toward the overall performance?

\item [RQ4:]Is \tool's performance on \dfj also reflected on \bugsjar?

\end{description}

\subsection{Training \tool}
For training \tool, we used all the one-line bug-fixes from the subjects in \bugsjar. For all these bugs, we extracted all the positive \rexps (that actually used in the repair) and all the negative \rexps with their feature vectors. Since negative \rexps are a lot more than the number of positive \rexps, to balance the training dataset, we replicated each positive \rexps 4 times, and for each positive \rexp we randomly chose equal number of negative but similar (e.g., variables or MIs) \rexps. Similar strategy has been used in other work as well~\cite{AML}. In this way, we obtained 1,580 data points, which is sufficient for our prediction model~\cite{sample:size}.

For evaluating \tool on \dfj, we used training dataset from all subjects except Commons Math in \bugsjar. However, for evaluating \tool on \bugsjar, we removed the subject under evaluation from the training dataset. This makes sure that our training and testing dataset are always mutually exclusive. We followed the standard 10-fold cross validation methodology to train \tool.

\subsection{Evaluation Metric and Patch Correctness} \label{sec:patch-correctness}

We evaluate each tool in terms of number of correct and incorrect patches.  
We classify a patch as \textit{correct}, if it is semantically equivalent to the developer-provided patch, based on a manual examination. This is consistent with previous work~\cite{Kali:ISSTA2015, SPR:FSE2015, Durieux:CoRR2015, Prophet:POPL2016, Angelix:ICSE2016}. An \textit{incorrect patch} is a patch that is not correct. To determine the correctness of a patch, two authors of the paper evaluated all the patches independently. In case of disagreement, we all had a group discussion until we had a mutual agreement.

\subsection{Experimental Configurations}

{\bf System.} We used 2 Core of Intel(R) Core(TM) i7-4790 CPU of 3.60GHz and 4GB memory per instance for our experiment. We used Ubuntu 14.04 LTS operating system and Java 7. 

{\bf \tool.} Currently, \tool iterates through top 200 statements returned by the bug localization tool. For each schema, \tool selects top 50 candidates returned by logistic regression model. \tool's  timeout is set to 90 minutes.

\section{Experimental Results}\label{results}

\subsection{RQ1: Comparison with state-of-the-art approaches}

To compare \tool with state-of-the-art, we chose five state-of-the-art G\&V repair tools: jGenProg~\cite{martinez2016automatic}, NOPOL\cite{martinez2016automatic}, a reimplementation of PAR (we call PAR$'$) by Le et al.~\cite{HDRepair:SANER2016}), history driven repair (we call HD-Repair)~\cite{HDRepair:SANER2016}, and ACS~\cite{ACS:ICSE2017}. To the best of our knowledge, these include all the repair tools that were evaluated on \dfj. Since automatic program repair experiments are very expensive, we discarded all the bugs that required multi-hunk fixes since they are by definition out of scope of \tool. Similar strategy is also followed by Le et al.~\cite{HDRepair:SANER2016} while evaluating HD-Repair. Therefore, the presented results of \tool for 82 bugs that required a single-hunk fix. The results of other tools are taken from the respective papers~\cite{martinez2016automatic, HDRepair:SANER2016, ACS:ICSE2017}.  Table~\ref{tbl:comparison} presents the results of each tool in terms of number of correct and incorrect patches. The overall results show that \tool produced 26 {\it correct} patches, which is the highest on \dfj. The second best is 18 patches by recently introduced ACS. All other tools generated 10 or less correct patches first. It should be noted that HD-Repair generated 16 patches but among them 10 were ranked first. Since \tool terminates at the first plausible patch, 10 is the fair number to compare for HD-Repair.

\begin{table}
\vspace{-1pt}
\setlength{\tabcolsep}{4pt}
\begin{center}
\caption{Comparison with existing techniques (Correct/Incorrect)}
\vspace{-5pt}
\label{tbl:comparison} 
\begin{tabular}{lccccc}
\toprule
{\bf Subject}&{\bf C.Math}&\bf {C.Lang}&{\bf Joda-Time}&{\bf JFreeChart}&{\bf Total}\\\midrule
\tool		& 12/7	&	8/4	&	2/1	&	4/3	&	{\bf 26/15}\\
ACS			&	12/4		&	3/1	&	1/0	&	2/0	&	18/5\\
HD-Repair	&6/NR	&	7/NR	&	1/NR	&	2/NR	&	16(10*)/NR\\
NOPOL		&1/20	&	3/4	&	0/1	&	1/5	&	5/30\\
PAR$'$		&2/NR		&	1/NR	&	0/NR	&	0/NR	&	3/NR\\
jGenProg		&5/13		&	0/0	&	0/7	&	0/2	&	5/22	\\
\bottomrule
\end{tabular}
\end{center} 
NR=Not Reported.\\
* HD-Repair generated correct patches for 16 defects, but only 10 were ranked first~\cite{ACS:ICSE2017}. All other tools terminate at the first plausible patch.
\vspace{-10pt}
\end{table}

Since ACS fixed the second highest number patches, we further investigated the nature of patches by ACS and \tool. We observed that only 4 patches are common between \tool and ACS. This observation matches our expectation since \tool targets more on MI-related bugs, whereas ACS targets condition-synthesis related bugs. It would be interesting to investigate how a tool performs that combines \tool and ACS. However, that is beyond the scope of our current evaluation. Although bug specific results are not available for HD-Repair, from the results it is clear that \tool (26 patches) fixed a lot more new bugs than HD-Repair (10 patches).

\begin{table}
\vspace{-1pt}
\setlength{\tabcolsep}{4pt}
\begin{center}
\caption{Contribution of Program Transformation Schemas (\tool)}
\vspace{-5pt}
\label{tbl:schema-contribution} 
\begin{tabular}{lcc}
\toprule
{\bf Transformation Schema}&{\bf Correct}&\bf {Incorrect}\\\midrule
Change in MI		&	12&6\\
Change in Boolean Expression	&6&8\\
Insertion of MI		&3&0\\
Type Widening			&2&0\\
Change in Return Expression		&2&0\\
If Guard (Null/Array Size Checking)  & 1&1\\
\bottomrule
\end{tabular}
\end{center} 
\vspace{-20pt}
\end{table}

We also investigated which schemas of \tool contributed more in  generating the plausible (both correct and incorrect) patches. From Table~\ref{tbl:schema-contribution}, we observe that changing MI in a comprehensive way is the most effective schema. It generated 12 correct patches, although it generated 6 incorrect patches. The second effective is the broad category of changes in Boolean expressions. This comprises several schema described in Section~\ref{lb:schemas}. They generated 6 correct patches. However, these schemas are also the source of many incorrect patches (8). Insertion of MI generated 3 correct patches with no incorrect patches. This result clearly demonstrates the effectiveness of \tool on (prevalent) MI-related bugs.

\subsection{RQ2: Contribution of \tool's \RExps and Ranking and Selection}

From the results of RQ1, it is hard to understand the sole contribution of our rich \rexps since various tools targeted various kinds of bugs, used various \rexps and transformation schemas. Therefore, we ran our two baselines: \baselineOne and \baselineTwo on the same set of bugs in \dfj that we used for RQ1. Recalling from Section~\ref{lbl:implementation}, both  baselines have the same program transformation schemas as of \tool. \baselineOne uses the \rexps following the existing tools, and \baselineTwo uses \tool's \rexps but selects  50 patches (same number as \tool) randomly instead of any machine learning technique. We ran \baselineOne 10 times due to its randomness, and counted the patches as correct even it generated the correct patches for one out of 10 times. 

\begin{table}
\vspace{-1pt}
\setlength{\tabcolsep}{4pt}
\begin{center}
\caption{Contribution of \tool's Ingredients, and Ranking and Selection of Candidate Patches}
\vspace{-5pt}
\label{tbl:comparison:baseline} 
\begin{tabular}{lcccc}
\toprule
{\bf Variant} & {\bf \RExps}& {\bf Ranking} & {\bf Correct} &	{\bf Incorrect}\\\midrule
\baselineOne&Traditional& Off	&	14	&	16\\
\baselineTwo	& Extended & Random	& 13&	5\\
\tool&	Extended & LR& 26	&	15\\
\bottomrule
\end{tabular}
\end{center} 
~~~~~LR = Logistic Regression
\vspace{-15pt}
\end{table}

Table~\ref{tbl:comparison:baseline} shows that even though we use the same  set of transformation schemas of \tool, \baselineOne can fix 14 bugs. By comparing these results with Table~\ref{tbl:comparison}, we see that \baselineOne is almost as good as ACS, and outperforms other tools. However, \baselineOne cannot generate any correct patches for the bugs that we presented in the motivating examples. Therefore, this result clearly demonstrates the contributions of our rich \rexps.

The results of \baselineTwo demonstrates that simply extending the set of \rexps without any effective selection and pruning actually decreases the number of correct patches (14 vs. 13). In our experiments, we observed that median size of our expanded \rexps is 30 times larger than the basic \rexps. Certainly we cannot apply and validate all the patches resulting from them. Therefore, we picked 50 patches randomly. When we investigated each correct patch from \baselineTwo, we observed that 10 patches are the same as that of \baselineOne. Six of these 10 patches are not affected by the expanded \rexps because the used schemas (T1-T5 in Section~\ref{lb:schemas}) are \rexps independent. Due to expanded \rexps, although \baselineTwo  generated correct patches for 3 new bugs that \baselineOne could not produce, it lost patches for 4 bugs that \baselineOne produced. However, with the machine learning technique, \tool has not lost any patches that \baselineOne generated, and additionally it generated correct patches for 12 more bugs.

\subsection{RQ3: Effect of Each Feature in Ranking and Selection }
To investigate whether all the features contributed in the ranking and selection of candidate patches, we turned off one of the four features at a time during the training and testing phase of \tool. The results in Table~\ref{tbl:feature:effect} indeed show that each feature  contributed in the ranking of correct patch in the search space. Turning off any feature reduces the number of correct patches. Among them distance turned out to be the least influential feature whereas the bug report and frequency are the dominant features. More specifically, \tool lost only one patch  in absence of distance, whereas lost 5 and 6 patches in absence of frequency and bug report respectively.

\begin{table}
\vspace{-1pt}
\setlength{\tabcolsep}{4pt}
\begin{center}
\caption{Effect of Each Feature in Ranking and Selection}
\vspace{-5pt}
\label{tbl:feature:effect} 
\begin{tabular}{lcc}
\toprule
{\bf Features} & {\bf Correct} &	{\bf Incorrect}\\\midrule
All-Distance	& 	25	&	15\\
All-Context	&	23	&	15\\
All-Bug Report	&	20	&	19\\
All-Frequency	&	21	&	17\\
\bottomrule
\end{tabular}
\end{center} 
\vspace{-15pt}
\end{table}

\begin{figure}[t]
\begin{Verbatim}[frame=single,fontsize=\footnotesize,commandchars=&\[\]]
   public static float max(final float a, final float b) {
&fvtextcolor[codered][-      return (a <= b) ? b : (Float.isNaN(a+b) ? Float.NaN : b);] 
&fvtextcolor[codeblue][+      return (a <= b) ? b : (Float.isNaN(a+b) ? Float.NaN : a);] 
    }
\end{Verbatim}
\vspace{-3mm}
\caption{Fix of MATH-482}
\vspace{-5mm}
\label{fig:math-482}
\end{figure}

\begin{figure}[t]
\begin{Verbatim}[frame=single,fontsize=\footnotesize,commandchars=&\[\]]
     if (escapingOn && c[start] == QUOTE) {
&fvtextcolor[codeblue][+        next(pos);]
         return appendTo == null ? null : appendTo.append(QUOTE);
     }
\end{Verbatim}
\vspace{-3mm}
\caption{Fix of LANG-477}
\vspace{-3mm}
\label{fig:lang-477}
\end{figure}

Figure~\ref{fig:math-482} presents an example where \tool could not generate the correct patch when we turned off the frequency feature. From the bug-fix we can see that repair location returns a very complicated expression which is basically a ternary {\tt if} expression. Here the mistake is that {\tt b} is returned instead of {\tt a}. This repair location generates a huge number of candidate patches since it has MI, return statement etc. In fact, each of {\tt a}, {\tt b}, and {\tt Float.NAN} can be replaced by 234 \rexps. Some of the candidates include {\tt Float.POSITIVE\_INFINITY}, {\tt Float.MIN\_VALUE}, {\tt min(a, b)} etc. However, since {\tt a} is one of the most frequent variables in the context, it got a high score and \tool generated the correct patch.

The example in Figure~\ref{fig:lang-477} shows the effect of bug report similarity feature. \tool synthesizes 944 valid MIs for the repair location. From the bug-fix, we see that the inserted MI does not have any element from the context. However, since the bug report contains both the terms {\tt next} and {\tt pos}, \tool ranks it at 3rd position. Due to lack of space, we do not provide concrete examples for distance and context features.

\subsection{RQ4: Effectiveness of \tool on \bugsjar}
Finally, we ran \tool on \bugsjar to understand its effectiveness on \bugsjar w.r.t. \baselineOne. Since \bugsjar has a large number of bugs, we randomly sampled 50\% of one-hunk bugs for this experiment. We could not run our tools on Log4J2 due to some engineering issues. More specifically, our framework and many of its dependent libraries used Log4J2 for logging, which was interfering with the ``subject Log4J2". This is an interesting engineering problem to solve in the future. Therefore, we randomly picked 127 bugs from 7 subjects in \bugsjar to create our sample set.

From Table~\ref{tab:elixir-on-bugsjar}, we observe that \tool generated correct patches for 22 bugs, whereas \baselineOne generated a correct patch for 14 bugs. The 8 new correct patches generated by \tool involved 4 MI insertions, 2 MI changes, 2 fields.  It should be noted that in RQ1, we demonstrated that \baselineOne is better than state-of-the-art techniques except ACS. Therefore, the results in Table~\ref{tab:elixir-on-bugsjar} demonstrate that \tool is similarly effective on \bugsjar compared to the state-of-the-art. The similarity in improvement also demonstrates that {\it \bugsjar is not biased toward \tool}.

\begin{table}
\footnotesize
\begin{center}
\caption{Patch Generation Summary (Correct/Incorrect) on \bugsjar}
\label{tab:elixir-on-bugsjar}
\vspace{0mm}
\begin{tabular}[h]{lccccc}
\toprule
{\bf Project} & {\bf In Sample} & {\bf \tool} & {\bf \baselineOne} \\ \midrule
Accumulo &  10 & 1/0 & 1/0\\
Camel &  16 &  2/1 & 1/0 \\
Commons Math &21 & 8/3 & 6/4 \\
Flink  & 7 & 2/0 & 1/0\\
Jackrabbit Oak &31 & 3/6 & 2/4  \\
Maven & 5 & 0/0 & 0/0 \\
Wicket  & 37 & 6/7 & 3/8 \\ \midrule
Total & 127 & 22/17 & 14/16\\
\bottomrule
\end{tabular}
\end{center}
\vspace{-5mm}
\end{table}

When we investigated the correct patches by \tool, we found many small but complicated fixes. Figure~\ref{fig:camel-7241} presents a concrete example that shows the fix for {\tt CAMEL-7241}, where \tool replaced a parameter of an MI, which is also an MI by a completely different MI that \tool synthesized.

\begin{figure}[t]
\begin{Verbatim}[frame=single,fontsize=\footnotesize,commandchars=&\[\]]
public static String toString(ByteBuffer buffer, 
    	  Exchange exchange) throws IOException {
&fvtextcolor[codered][-   return IOConverter.toString(buffer.array(),exchange);] 
&fvtextcolor[codeblue][+   return IOConverter.toString(toByteArray(buffer),exchange);]
}
\end{Verbatim}
\vspace{-3mm}
\caption{Fix of CAMEL-7241}
\vspace{-5mm}
\label{fig:camel-7241}
\end{figure}

\section{Threats to Validity}\label{sec:threats}

\textbf{External validity.}
Our evaluation is conducted only on the \dfj and \bugsjar datasets and our conclusions may not generalize to subject systems and bugs beyond these datasets. The use of two, independently constructed datasets was in part a conscious choice to mitigate this threat. Further, in constructing \bugsjar we followed a rigorous, scientific procedure to identify 8 diverse, representative subject systems, and included \textit{all} their reproducible, single-module bugs in our dataset. Second, \tool has currently only been instantiated and validated on Java application bugs. While the \tool technique is conceptually general, the current results may not generalize to bugs in other OO-languages, such as C++.

\textbf{Internal validity.} 
Any implementation error of \tool could impact internal validity. We mitigated this by manual inspection of all patches produced by the tools and by following good development and QA practices. 

The \bugsjar dataset and hence our use of it may pose several threats too. In creating \bugsjar, we followed a rigorous and scientific procedure for selecting subject systems. Therefore, \bugsjar is not biased toward any specific type of bugs or subjects that is beneficial to \tool.
Further, errors in the classification of commits as bugs or feature-enhancements, due to mis-labeling in JIRA, or errors in our bug-extraction scripts, could pose a threat. We mitigated this threat by manually re-examining each selected bug to ensure it was indeed a bug. Flaky bugs, which may produce unpredictable behavior during repair, also pose a threat, which we mitigated by selecting only bugs that consistently re-produced in 10 runs.

Finally, \baselineTwo{}, which uses randomized prioritization of candidate patches, is run 10 times to generate a patch.

\textbf{Construct validity.} Our criterion for classifying patches as \textit{correct} or \textit{incorrect} is based on manual analysis, which is not scientifically rigorous, even though it is accepted practice in previous work~\cite{Kali:ISSTA2015, SPR:FSE2015, Durieux:CoRR2015, Angelix:ICSE2016}. We tried to mitigate this threat by following the protocol in Section~\ref{sec:patch-correctness}, which involved independent classification of the patches by the first two authors and reconciliation of any discrepancies through deeper examination of the patch in question by all authors.

\section{Related Work}\label{related_work}

\textbf{Search-based repair for C programs.} GenProg~\cite{GenProg:ICSE2012}, which pioneered this area, uses genetic programming to search a space of repair mutations formed by code snippets copied from elsewhere in the program. RSRepair~\cite{RSRepair:ICSE2014} uses random search instead, while AE~\cite{AE:ASE2013} uses deterministic search coupled by analysis to prune equivalent patches. Relifix~\cite{Relifix:ICSE2015} proposes a set of specialized repair schemas customized for repairing software regression errors. SPR~\cite{SPR:FSE2015} prioritizes repair of conditional statements, using abstract repair conditions to implicitly evaluate and prune away infeasible condition repair candidates (staging) before generating concrete repairs.  In recent work, Tan et al.~\cite{antiPatterns:FSE2016} propose the use of \textit{anti-patterns} -- a set of generic forbidden repair transformations, to reduce the incidence of plausible patches, that typically arise in search-based program repair, due to weak test-suites. The above contributions can, and in part have, been re-purposed for repair of Java (or other OO) programs. Our proposed technique seeks to substantially expand the repair space, by using richer repair expressions incorporating method invocations, and by efficiently searching this space using a machine-learnt model. Thus, it nicely complements the above body of work. 

\textbf{Search-based repair for Java programs.} PAR~\cite{PAR:ICSE2013} overlays GenProg's search strategy with a set of repair templates manually derived from human-written patches. History-driven repair~\cite{HDRepair:SANER2016} uses a rich set of templates drawn from GenProg, PAR, and mutation testing, to produce a large pool of candidate repairs which it then prioritizes and prunes based on the frequency of previous (human-written) patches. In a very recent work, ACS~\cite{ACS:ICSE2017} proposes a method for precise condition synthesis by instantiating variables in predicates that frequently occur in a given corpus of code, using various heuristics to rank and choose the variables. While each of these techniques handles the Java language as such, unlike us, they use method invocations in very limited and specific ways, ostensibly to avoid an explosion in the repair search space. By contrast, the expanded and generalized use of method invocations in repairs is the main contribution of our work.

\textbf{Oracle-based repair.} SemFix~\cite{SemFix:ICSE2013} uses symbolic execution to create an oracular representation of an expression under repair and then uses program synthesis to generate a repaired statement compatible with this ``oracle". MintHint~\cite{MintHint:ICSE2014} follows SemFix in creating the oracle but uses statistical analysis to search for a repair. DirectFix~\cite{DirectFix:ICSE2015} generates \textit{minimal} repairs to obtain comprehensible repairs by encoding the problem as a partial maximum satisfiability problem over SMT formulas. Angelix~\cite{Angelix:ICSE2016} solves the scalability problems of DirectFix by using a lightweight repair constraint. SearchRepair~\cite{SearchRepair:ASE2015} uses \textit{semantic search} to search a corpus of human-written patches, encoded as satisfiability modulo theories (SMT) constraints, for possible matches to a repair problem. All of these techniques target C programs. The only exception is NOPOL~\cite{NOPOL:TSE2017}, which targets Java programs. It focuses on the repair of branch conditions, using instrumented test-suite executions to synthesize an oracle, which is converted into a suitable SMT formula and solved to obtain a patch. Like search-based Java repair techniques, NOPOL also incorporates method invocations in a very limited way to curb explosion of the repair space. In principle, our repair space ranking ideas could be applied in the patch-synthesis stage of an oracle-based repair technique, \ie in generation of a concrete patch from the oracle. This could constitute interesting future work. 

\textbf{AI in program repair.} This is a nascent branch of research in program repair. Prophet~\cite{Prophet:POPL2016} builds on the SPR technique, further using a machine learned model of previously-known correct human patches to prioritize candidate repairs. Conceptually, \tool\ also uses machine learning to rank repair candidates. However, unlike Prophet's elaborate model with over 3000 features, which produces only a $25\%$ improvement in repair outcomes (from 12 to 15 patches) we use a simple model with only 4 potent features, interestingly with substantially better repair outcomes ($85\%$ improvement from 14 to 26 patches). Also, our use of machine learning is paired with a meaningful expansion in the repair space to target a specific aspect of patches, \ie incorporating MIs in repair expressions. DeepFix~\cite{DeepFix:AAAI17} employs deep learning to fix language-level common programming errors in C programs. This work while interesting in its own right, is somewhat different from the test-suite based repair of functional errors, targeted by vast majority of program repair research, including ours.

\vspace{-5pt}
\section{Conclusions}\label{conclusions}
\vspace{-5pt}
This work was motivated by the extensive use of method invocations (MI) in object-oriented (OO) programs, and indeed their prevalence in patches of OO-program bugs. We proposed a generate-and-validate repair technique, \tool, designed for repair of OO-programs, and instantiated it for Java program repair. \tool\ aggressively uses MIs, on par with local variables, fields, and constants, to construct more expressive repair expressions, that go into synthesizing patches. The ensuing enlargement of the repair space is effectively tackled by using a machine-learnt model to rank concrete repairs. The model relies on four features derived from the program context, \ie the code surrounding the repair location, and from the bug report. \tool\ was evaluated on two separate datasets, namely the popular \dfj dataset, and another large-scale dataset \bugsjar created by us. The evaluation shows that, by enlarging and effectively searching the larger repair space, \tool is able to significantly increase the number of correctly repaired bugs 
while also out-performing other state-of-the-art tools like ACS, HD-Repair, NOPOL, PAR, and jGenProg.

We believe that this work is a promising demonstration of how AI/ML techniques can be used to expand the scope of automatic program repair techniques. In future, we hope to pursue and realize the full potential of this branch of research.

\balance

\bibliographystyle{IEEEtran}
\bibliography{references}

\end{document}